# Diamond Nitrogen-Vacancy Center Magnetometry: Advances and Challenges

Jixing Zhang[1], Lixia Xu[1], Guodong Bian[1], Pengcheng Fan[1], Mingxin Li[1], Wuming, Liu[2], and Heng Yuan[1,3,4,5]

[1]School of Instrumentation and Optoelectronic Engineering, Beihang University, Beijing 1000191, China;
[2]Beijing National Laboratory for Condensed Matter Physics, Institute of Physics, Chinese Academy of Sciences, Beijing 100190, China
[3]Research Institute of Frontier Science，Beihang University, Beijing 1000191, China;
[4]Beijing Advanced Innovation Center for Big Data-Based Precision Medicine, Beihang University, Beijing 100191, China;
[5]Beijing Academy of Quantum Information Sciences, Beijing 100193, China;

**Abstract**

Diamond nitrogen-vacancy (NV) center magnetometry has recently received considerable interest from researchers in the fields of applied physics and sensors. The purpose of this review is to analyze the principle, sensitivity, technical development potential, and application prospect of the diamond NV center magnetometry. This review briefly introduces the physical characteristics of NV centers, summarizes basic principles of the NV center magnetometer, analyzes the theoretical sensitivity, and discusses the impact of technical noise on the NV center magnetometer. Furthermore, the most critical technologies that affect the performance of the NV center magnetometer are described: diamond sample preparation, microwave manipulation, fluorescence collection, and laser excitation. The theoretical and technical crucial problems, potential solutions and research technical route are discussed. In addition, this review discusses the influence of technical noise under the conventional technical conditions and the actual sensitivity which is determined by the theoretical sensitivity and the technical noise. It is envisaged that the sensitivity that can be achieved through an optimized design is in the order of 10 fT·Hz$^{-1/2}$. Finally, the roadmap of applications of the diamond NV center magnetometer are presented.

# 1 Introduction

Nitrogen-vacancy (NV) center is a one of hundreds of different color centers in diamonds [1] and has represented a hot topic of research as its unpair electron spin is feasible to achieve polarization, manipulation, and readout. NV center has been extensively used in several fields, including quantum registers [2], quantum computing [3], time measurement [4], electric field measurement [5], rotation sensing[6], and nuclear magnetic resonance applications [7]. Among these, magnetometry is one of the most important applications. However, important issues regarding the diamond NV center magnetometer still need to be addressed. Some of the key questions surrounding this very active research topic, which will be addressed in this review, are: What are the advantages and unique features of the NV center magnetometer compared to other existing magnetometers? How great is its sensitivity potential? What are its potential application scenarios? The answers to these questions will determine the possibility to implement the NV center magnetometer as a practical technology rather than a short-lived research topic. The purposes of writing this review are to provide a thorough summary of the NV center magnetometer and to initiate a meaningful discussion of the future research directions for this technique.

Magnetometers have a wide range of applications in engineering, science, and medicine [8][9]. The mainstream types of magnetometer, are introduced as follow [10][11][12][13]. **Hall magnetometer.** Base on Hall effect, the accuracy of the Hall magnetometer is poor, against small size and low cost. and it is widely used in various scenarios. **Magnetometer with magnetoresistance (MR) effect.** There are several types of these magnetometers, such as the giant magnetoresistance (GMR), the anisotropic magnetoresistance (AMR), and tunneling magnetoresistance (TMR) magnetometers. MR Magnetometers based have also been extensively studied and represent still a very hot research field [14][15]. **Inductive pickup coils.** The accuracy of the inductive sensor [16] can vary greatly. This kind of magnetometer is used to measures AC magnetic fields and can reach a maximum sensitivity of 20 fT·Hz$^{-1/2}$. **Fluxgate magnetometer.** This magnetometer is composed of a core with high magnetic flux, an excitation coil, and a detection coil. It is a commonly used pT-level magnetometer with a good comprehensive performance. **SQUID magnetometer.** The superconducting quantum interference device (SQUID) magnetometer is based on the e superconducting effect[17]. It has been widely used [18] and it remains one of the most sensitive magnetometers. **Atomic magnetometer.** This type of magnetometer and the NV center magnetometer are based on the electronic spin[19]. They include the optical pump magnetometer, the laser optical pump magnetometer, the Overhauser magnetometer, the proton magnetometer, and the spin exchange relaxation-free (SERF) magnetometer, which can achieve a sensitivity in the order of the fT·Hz$^{-1/2}$. **Fiber-based magnetometer.** Thanks to the use of ultra-high Verdet constants, the optical rotation effect employed in the total optical fiber-based magnetometer can reach a sensitivity of 20 fT·Hz$^{-1/2}$ [20]. Fiber-based Magnetometers using magnetic fluids have also been studied [21][22].

The typical sensitivity and measurement range of the magnetometers listed above are illustrated in Fig. 1. It can be noticed that the NV center magnetometer can surpass the fluxgate magnetometer performance, and has the potential to exceed the sensitivity of the optical pump magnetometer in the future. Furthermore, it was pointed out that compared with other magnetometers, the diamond NV center magnetometer has the potential to approach the theoretical limit of magnetic field measurements[23].

This article is composed of the following main four parts: **1.** The basic physical properties of the NV center are discussed. These include its energy level structure, optical dynamic properties, spin Hamiltonian and spin dynamics, charge-state conversion, nuclear spin ultra-fine energy levels, and the effects of temperature and stress. **2.** Four methods of DC magnetic measurements are analyzed. Furthermore, the influence of technical noise on the magnetic measurement results and transfer

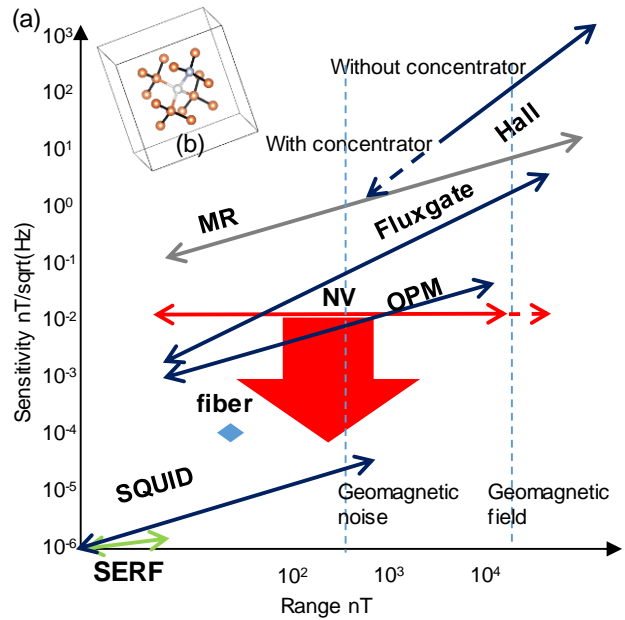

**Fig. 1 (a)** Sensitivity versus measurement range for different magnetometry technologies. SERF magnetometer and SQUID magnetometer have the highest sensitivity. The current sensitivity of NV center magnetometer is slightly worse than fluxgate magnetometer and optical pump magnetometer but higher than MR magnetometer and Hall magnetometer. The end of the red arrow reaches the sensitivity limit of the NV center magnetometer under current technologies, which is about 10 fT/Hz$^{1/2}$. **(b)** NV defect in diamond. White ball is vacancy, gray ball is nitrogen atom, and orange balls are carbon atoms.

coefficient are discussed. The principle of AC magnetic field measurements is briefly introduced. **3.** The technical solutions of the four key technologies of the NV center magnetometer are presented, alongside the technical indexes that can be achieved. Several special DC magnetic field measurement schemes and non-fluorescent spin readout schemes are introduced, and the magnitude of the technical noise under the existing technical conditions is analyzed. The actual sensitivity is evaluated. **4.** The overall scheme and corresponding main application of the NV magnetometer are illustrated.

# 2 NV physical characteristics

The NV center characteristics have been extensively studied [24]. In the NV center, a substitute nitrogen atom is connected to a vacancy to form the NV center, which contains five unpaired electrons as the $NV^0$ center. If an extra electron is captured, it becomes the $NV^-$ center. In this article, unless otherwise specified, NV refers to $NV^-$.

## 2.1 Energy level structure and optical dynamics of the NV center

### 2.1.1 Energy level structure and Intrinsic optical dynamics

The NV center satisfies the symmetry of $C_3V$. Based on six electrons model and $C_3V$ symmetry, the ground state triplet state, singlet states, excited state of NV center has been fully established [25] [26][27][28], as shown in Fig. 2. At room temperature, a 532 nm laser is generally used to excite the ground state into the excited phonon sideband. This process is spin-maintained, and the pumping rate $k_{24} = k_{13} = \Gamma$ is proportional to the optical power density. The excited state spontaneously relaxes back to the ground state, which is also spin-maintained, with a pumping rate $k_{42} = k_{31}$. During this process a photon of 637-800 nm is generated. The transition between singlet and excited state was originally prohibited, but due to the inter-system crossing (ISC) effect between state 5 and the excited state, the excited state can relax spontaneously to singlet [29]. Furthermore, the process is spin-related, with the relaxation rate $k_{45}$ of state 4 being much greater than the relaxation rate $k_{35}$ of state 3. The relaxation rate between singlet 5 and singlet 6 is $k_{56}$. In line with recent theoretical studies, the relaxation from singlet to ground state can also be explained theoretically [30].

Due to the ISC effect, the relaxation rate $k_{61}$ for the single state to ground state 1 is faster than the relaxation rate $k_{62}$ to ground state 2. The complete transition process is shown in Fig. 2. As a result of the above two processes, it can be easily concluded that under the action of an excitation light, the optical loop makes the distribution of its ground state $m_s = 0$ higher than the equilibrium state, i.e. polarized to $m_s = 0$. Furthermore, since the relaxation rate $k_{45}$ is higher than $k_{35}$, the photon emission probability in state 4 is less than in state 3. This results in the fluorescence intensity being related to the spin state, thus optical spin detection is realized. $\mathcal{F}$ represents the number of fluorescent photons detected per unit time. And $I = \int_0^{T_R} \mathcal{F}(t)dt$ represents the total number of fluorescent photons detected during a laser pulse. In general, the difference between the $m_s = 0$ related fluorescence $I_0$ and the $m_s = \pm 1$ related fluorescence $I_1$ for the single center is $I_0 - I_1 = 30\%$, while it is about 10% for an ensemble[31]. The fact that $k_{45}$ is higher than $k_{35}$ also leads to the relation between population of singlet state. Thus, the 1042 nm spin-singlet transition also could achieve spin readout.

### 2.1.2 Extrinsic optical dynamics

For quantum sensing applications $NV^-$ is used rather than $NV^0$. However, under certain conditions, a transition can occur between $NV^-$ and $NV^0$. The stimulated transition between $NV^-$ and $NV^0$ can be achieved by a two-photon process or a one-photon process, as shown in Fig. 2. Since $NV^0$ does not have optical polarization characteristics, it is not easily manipulated, and it is benefit to increase the proportion of $NV^-$[32]. Furthermore, the single state does not participate in the ionization process, thus the charge state conversion is also related to the spin state[33][34][35]. The spontaneous transitions between $NV^-$ and $NV^0$ might in the dark have also been measured[36].

According to the above analysis, it is possible to derive the master equation, including the internal and external optical dynamics, as follows:

$$\begin{aligned}
\dot{\rho}_{11} &= -\Gamma\rho_{11} + k_{31}\rho_{33} + k_{51}\rho_{55} \\
\dot{\rho}_{22} &= -\Gamma\rho_{22} + k_{42}\rho_{44} + k_{52}\rho_{55} \\
\dot{\rho}_{33} &= \Gamma\rho_{11} - (k_{36} + k_{31} + k_{37})\rho_{33} + k_{73}\rho_{77} \\
\dot{\rho}_{44} &= \Gamma\rho_{22} - (k_{46} + k_{42} + k_{47})\rho_{44} + k_{74}\rho_{77} \\
\dot{\rho}_{55} &= k_{65}\rho_{66} - (k_{51} + k_{52} - k_{56})\rho_{55} \\
\dot{\rho}_{66} &= k_{36}\rho_{33} + k_{46}\rho_{44} + k_{56}\rho_{55} - k_{65}\rho_{66} \\
\dot{\rho}_{77} &= -(k_{73} + k_{74} + k_{78})\rho_{77} + k_{37}\rho_{33} + k_{47}\rho_{44} + k_{87}\rho_{88} \\
\dot{\rho}_{88} &= -(k_{81} + k_{82} + k_{87})\rho_{88} + k_{18}\rho_{11} + k_{28}\rho_{22} + k_{78}\rho_{77}
\end{aligned} \quad (1)$$

where $\rho_{ii}$ represents the $i$th diagonal term of the density matrix and the population of $i$th state. The dynamics determined by Eq. (1) is shown in Fig. 2.

## 2.2 Ground state spin Hamiltonian and spin dynamics

Since only the ground state of the NV color center is stable, the magnetic measurement uses the relationship between the ground state spin dynamics and the magnetic field. The spin dynamics of the ground state of the NV center can be described by the Hamiltonian as follows:

$$H = D_{gs}S^2 + \mu_e\vec{B}\cdot\vec{S} + A\vec{S}\cdot\vec{I} + QI^2 + \left(\mu_N\vec{B}\cdot\vec{I} + \sum A_c\vec{S}\cdot\vec{I_c} + \cdots\right) \quad (2)$$

The first term in Eq. (2) is the zero-field splitting term, along the direction of the NV axis, while the second term is the Zeeman splitting term, in which the gyromagnetic ratio of the electron spin $\mu_e$ = 2.8 MHz/G. Optical detection magnetic resonance (ODMR) measures the energy level shift caused by this item, so it could be used as the basic principle of magnetic measurement[37].

The third term in Eq. (2) is the hyperfine interaction between host nitrogen nuclear spin and electrons, $A$ is the hyperfine interaction tensor, with $A_\parallel$ = 2.1 MHz and $A_\perp$ = 2.2 MHz. Under natural conditions, the abundance of the isotope $^{14}$N is 99.64%. The nuclear spin quantum number for $^{14}$N is 1. Through artificial preparation, also the $^{15}$N atomic isotope can be obtained. The nuclear spin quantum number in this case is 1/2. The fourth term is the nuclear spin electric quadrupole moment term, $Q$ = 5

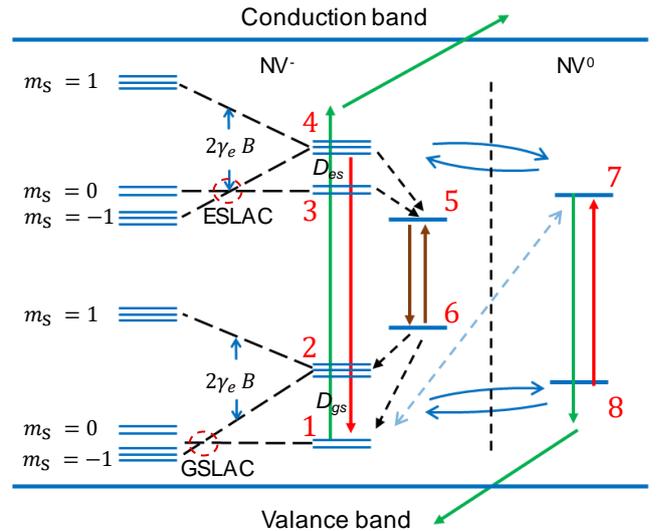

**Fig. 2** $NV^-$ and $NV^0$ energy level structures and transitions between energy levels. For $NV^-$ The excited state contains two triplet states. These energy levels can be clearly resolved at low temperature. However, at room temperature, the energy level of the excited state is subject to an averaging effect, thus the excited state can be considered as one triplet state. A zero-field splitting $D_{gs}$ = 2.87 GHz occurs between the ground state spin $m_s = 0$ (here referred to as state 1) and $m_s = \pm 1$ (here referred to as state 2). There is also a zero-field splitting $D_{es}$ = 1.42 GHz for the excited state with $m_s = 0$ (here referred to as state 3) and $m_s = \pm 1$ (here referred to as state 4). The zero phonon line (ZPL) between the ground state and the excited state is 1.945 eV (637 nm). The single state is mainly involved in a high-energy level 1A (here referred to as state 5) and a low-energy level 1E (here referred to as state 6), which have an energy level difference of 1.190 eV (1043 nm). ZPL between ground state 7 and excited state 8 of $NV^0$ is 2.156 eV(575 nm). $NV^-$ in the ground state absorbs a photon to reach the excited state, and then absorbs another photon to reach the conduction band, thus the electron energy exceeds its binding energy and becomes a free electron, and $NV^-$ becomes $NV^0$. Similarly, $NV^0$ can absorb two photons and emit a free hole, then becomes $NV^-$.

MHz. In parentheses, the Zeeman splitting term of nuclear spins, in which the gyromagnetic ratio of N nuclear spins is $\mu_N$ = 270 Hz/Gauss, and the hyperfine interaction of $^{13}$C nuclear spins are ignored in magnetic measurement. Considering four different axial directions of diamond and a magnetic field of arbitrary orientation, the magnetic field projection onto the four quantum principal axes may be different. Therefore, the energy level splittings are different, as shown in Fig. 9 (g). If the transverse magnetic field is small (< 100 G), the $S_{xy}$ term can be safely neglected as $D_{gs}$ is large. Below, a few key points are listed:

### 2.2.1 Hyperfine interaction term

When the transverse magnetic field is modest, the $^{14}$N nuclear spin processions along the NV axis, as the $^{14}$N electric quadrupole moment is very large. If the nuclear spin is not polarized, the effect of the $^{14}$N on the spin of the NV center electron can be considered as equivalent to that of a constant magnetic field [38]. $D_{gs}$ and $D_{es}$ lead to the energy level crossing of states 0 and 1 under the applied magnetic field of 500 G and 1000 G, which are named ground state level anticrossing (GSLAC) and excited state level anticrossing (ESLAC) [39][40][41][42] and lead to nuclear spin polarization, as shown in Fig. 2. For the $^{13}$C, its ultra-fine energy level is related to the relative position between the $^{13}$C and the NV center [43].

### 2.2.2 Temperature and strain term

$D_{gs}$ shifts with temperature. The offset is −74.2 (7) kHz/K [44]. This temperature-induced offset is considered to be local thermal expansion, which represents the change in lattice parameters caused by thermal expansion [45]. Factors for the electron-phonon coupling are also presented [46]. Various schemes for temperature measurements using NV centers have been proposed, which can achieve a measurement sensitivity in the order of the mK [47][48][49][50][51].Several schemes exist that employ mechanical vibrations to drive spins [52], such as the direct contact vibration method. It is possible to obtain the transition between $m_s = 1$ and $m_s = -1$ [53][54][55][56][57]. Another possibility relies on combining a magnetic probe, whereby the vibrations that drive the magnetic probe are used to generate a vibrating magnetic field, which then couples with the NV center [58][59][60]. At the same time, considering the unique advantages of diamond hardness, diamond NV center magnetometry measurements can be performed under extremely high pressure [61][62][63].

### 2.2.3 Decoherence time

The longitudinal relaxation time $T_1$ determined by the phonon vibrations inside the solid, thus it is related to the temperature. In room temperature, $T_1$ is order of ms. The coherence time $T_2$ reaches the order of 100 μs. the dephasing time $T_2^*$ can only reach the order of 1 μs. $2T_1$ is the upper limit of $T_2$, and $T_2$ is the upper limit of $T_2^*$. At liquid helium temperature (3.7 K), the decoherence time of a single NV center $T_1$ reaches the order of 1000 s, and a decoherence time of 1.58 s [64]. $T_2^*$ and $T_2$ are closely related to the diamond production process, so we will analyze them in detail at Section 4.1.

### 2.2.4 Spin dynamics

Considering the manipulation between ground substates by microwave (MW) with a single frequency $f$, the dynamics of the electron spin density matrix can be obtained using the rabi oscillation model.

$$\begin{aligned}
\dot{\rho}_{12} &= \left(-\frac{1}{T_2^*} + i\delta\right)\rho_{12} + i\frac{\Omega}{2}(\rho_{11} - \rho_{22}) \\
\dot{\rho}_{21} &= \left(-\frac{1}{T_2^*} + i\delta\right)\rho_{21} + i\frac{\Omega}{2}(\rho_{22} - \rho_{11}) \\
\dot{\rho}_{11} &= i\frac{\Omega}{2}(\rho_{12} - \rho_{21}) - \frac{1}{T_1}(\rho_{11} - \rho_{22}) \\
\dot{\rho}_{22} &= i\frac{\Omega}{2}(\rho_{21} - \rho_{12}) - \frac{1}{T_1}(\rho_{22} - \rho_{11})
\end{aligned} \quad (3)$$

where $\Omega$ is Rabi frequency which represents the MW power, $\rho_{ij}$ is off-diagonal term of the density matrix, and detuning is $\delta = f - (\mu_e B + D_{gs})$. In this way, combining the optical dynamic formula (Eq. 1) and the spin dynamic formula (Eq. 3), the theoretical basis of the master equation method for the magnetic measurement is established.

# 3 Principle and scheme of the magnetic measurement

In this section, the principles of four mainstream magnetic field measurements schemes are introduced and compared, including the Ramsey scheme, the pulsed ODMR (PUODMR) scheme, and the continuous wave ODMR (CWODMR) scheme [65]. CWODMR has an extending scheme, in which a lock-in (LI) amplifier is used for lock-in detection and it is abbreviated as CWLI.

## 3.1 Measuring principle and theoretical sensitivity

### 3.1.1 The Ramsey scheme

The measurement principle of the Ramsey scheme has been discussed in detail[32]. As shown in Fig. 3 (a), After the Ramsey sequence, the proportion of the |1> state is:

$$\rho_{11} = \frac{\Omega^2}{\Omega^2 + \delta^2}\left(\cos\frac{\delta T_S}{2} - \frac{\delta}{\sqrt{\Omega^2 + \delta^2}}\sin\frac{\delta T_S}{2}\right)^2 \quad (4)$$

When $\Omega$ is sufficiently large compared to the detuning frequency $\delta$, it can be regarded as: $\frac{\Omega^2}{\Omega^2+\delta^2} \to 1$ and $\frac{\delta}{\sqrt{\Omega^2+\delta^2}} \to 0$, resulting in $\rho_{11} = \frac{1-\cos\delta T_S}{2}$. Assuming that the change of magnetic field is $\Delta B$, the change of $\rho_{11}$ is

$$\Delta\rho_{11} = \frac{1}{2}\sin(\delta T_S) T_S \mu_e \Delta B \quad (5)$$

The scale factor between $\rho_{11}$ and the magnetic field is defined as:

$$k_{Ramsey} = \frac{\Delta\rho_{11}}{\Delta B} = \frac{1}{2}\sin(\delta T_S)\mu_e T_S \quad (6)$$

Due to spin dephasing, the Ramsey signal decays with time. The phenomenological attenuation term $e^{-pT/T_2^*}$ is added, with $p$ in the range of 1-2. Thus, the optimal $T_S$ and $\delta$, which maximize $k_{Ramsey}$, satisfy:

$$\begin{aligned} \delta T_S &= \frac{\pi}{2} + j\pi, j = 0, \pm 1, \pm 2, \cdots \\ T_S &= T_2^* \end{aligned} \quad (7)$$

With $\Delta I/\Delta m = (I_0 - I_1)$, the sensitivity formula can be expressed as:

$$\eta_{Ramsey} = \frac{\sqrt{T_P + T_2^* + T_R}}{k_{ramsey}F\sqrt{N}} = \frac{\sqrt{T_P + T_2^* + T_R}}{\pi T_2^*\gamma_e e^{-p}F\sqrt{N}} \quad (8)$$

where $F = \frac{1}{\sqrt{1+\frac{2(I_0+I_1)}{(I_0-I_1)^2}}}$ is the readout fidelity and $N$ is the number of NV centers involved in the sensing. Considering the low contrast, the following approximation is valid,

$$F\sqrt{N} \approx \frac{I_0 - I_1}{I_0 + I_1}\sqrt{I_{avg}} = C\sqrt{I_{avg}} \quad (9)$$

where, $C$ is the contrast.

### 3.1.2 Pulsed ODMR scheme

The PUODMR sequence is shown in Fig. 3 (b). In order to maximize the ODMR contrast, the pulse time is set to be π/2. In the PUODMR scheme, the relationship between fluorescence and magnetic field is phenomenologically defined as:

$$\Delta I_{pulse} = I_0 + (I_0 - I_1)\rho_{11}(\delta + \mu_e \Delta B, \Omega, T_2^*, T_1) \quad (10)$$

Both the ODMR scheme and the Ramsey scheme can be split into different steps: polarization, sensitivity, and readout. The readout of the two schemes are consistent and both satisfy Eq. (8-9). The scale factor is defined as: $k_{PUODMR} = \frac{\Delta\rho_{11}}{\Delta B}$. Thus, the sensitivity becomes:

$$\eta_{PUODMR} \approx \frac{\sqrt{T_I + T_2^* + T_R}}{k_{PUODMR}C\sqrt{I_{avg}}} \quad (11)$$

The largest $k_{PUODMR}$ is for optimal sensitivity. As ODMR approximately stratifies the Lorentz distribution, the largest $k_{PUODMR}$ for optimal sensitivity occurs at $\delta = \Delta\nu/\sqrt{3}$ and is:

$$k_{PUODMR} = \frac{4\Delta\nu}{3\sqrt{3}\mu_e} \quad (12)$$

where, $\Delta\nu$ is the half-width at half-maximum (HWHM) of OMDR, the limit of which is $\pi/T_2^*$.the microwave power $\Omega$ has influence on both $\Delta\nu$ and $C$, as will be discussed via simulations in Section 3.2.1.

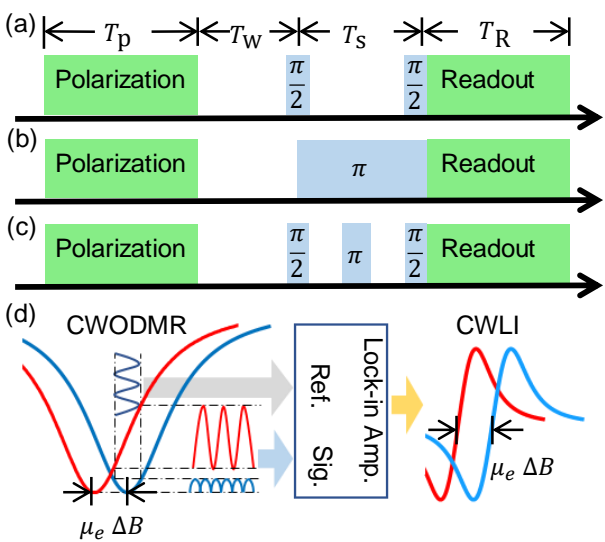

**Fig. 3** Protocols of four conventional schemes. Laser is depicted as green, while microwave is depicted as light blue. $T_P$ is the laser polarization pulse, and $T_R$ is the laser readout pulse. $T_W$ is the waiting time after polarization, during which the singlet population relaxes to ground state. $T_S$ is the free precession time under the magnetic field that has to be measured. **(a)** The Ramsey magnetic measurement sequence. A pulse of laser polarizes the spin $m_s = 0$, then a π/2 microwave pulse flips the spin state to |0> + |1>. After the free precession of time $T_S$, the spin state accumulates a phase related to the magnetic field. Assuming that the change from magnetic field is $\Delta B$, the corresponding phase is: $\phi = \Delta B \mu_e T_S$. When the second π/2 microwave pulse is sent, the phase change is modified into a spin state distribution change, and finally the spin is read out using a laser pulse. **(b)** PUODMR magnetic measurement sequence. **(c)** Spin echo sequence for AC measurement. **(d)** The CWODMR method measures the change of the magnetic field by directly readout of the shift of the ODMR spectrum (from the blue line to the red line). Through the lock-in amplifier, CWLI method readouts the derivative signal of ODMR by modulating the microwave

### 3.1.3 Continuous wave ODMR scheme

In CWODMR scheme, both microwave and the laser are continuously applied simultaneously. Under laser excitation, the total number of photons generated per unit time is defined as,

$$\mathcal{F}(\delta + \mu_e \Delta B, \Omega(t), \Gamma(t), T_2^*, T_1) \quad (13)$$

As the CWODMR has no separate steps, its scale factor is defined as $k_{CWODMR} = \frac{\Delta \mathcal{F}}{\Delta B}$. Similar to PUODMR scheme, at $\delta = \Delta \nu / \sqrt{3}$, $k_{CWODMR}$ has maximum,

$$k_{PUODMR} = \frac{3\sqrt{3} C_{cw} \mathcal{F}}{4\Delta \nu} \quad (14)$$

Due to the characteristics of the Poisson distribution, the standard deviation of $\mathcal{F}$ is $\Delta \mathcal{F} = \sqrt{\mathcal{F}}$. Thus, the sensitivity is expressed as:

$$\eta_{CWUL} = \frac{4\Delta \nu}{3\sqrt{3} C_{cw} \gamma_e \sqrt{\mathcal{F}}} \quad (15)$$

The optimal $\Omega$ and $\Gamma$ are verified via simulations in section 3.2.1.

### 3.1.4 Lock-in scheme

For CWODMR magnetic measurements, the adoption of a direct detection scheme would be severely affected by the low-frequency technical noise. Therefore, a more practical solution is provided by a phase-sensitive detection scheme based on a lock-in amplifier [66][67]. The frequency of the sensitive noise can be shifted from lower to higher values through the lock-in amplifier, resulting in the ability to effectively suppress the low frequency noise, as shown in Fig. 3 (d). By using the theoretical analysis method of the CWLI scheme in Ref [68], the theoretical scheme is calculated as follows. The microwave frequency modulates with modulation depth $\Delta f$ and modulation frequency $\omega_m$,

$$\delta(t) = \delta_0 + \mu_e \Delta B + \Delta f \sin(\omega_m t) \quad (16)$$

The lock-in amplifier requires a multiplier and a low-pass filter. An integration time equal to a modulation period is used to simulate low-pass filtering. Following DC isolator, multiplier, and filtering, the output is:

$$S_{cw}(B) = \frac{\omega_m}{2\pi} \int_{-\frac{\pi}{\omega_m}}^{\frac{\pi}{\omega_m}} \mathcal{F}(\delta_0 + \mu_e \Delta B + \Delta f \sin(\omega_m t)) \sin(\omega_m t) dt \quad (17)$$

The optimal conditions are verified via simulations in section 3.2.1.

### 3.1.5 AC magnetic field measurement principle

As previously discussed, the dephasing time determines the sensitivity of the magnetometer. It is therefore critical to study how to improve the dephasing time. Since the decoherence time $T_2$ is much larger than $T_2^*$, to increase the time limiting the sensitivity from $T_2^*$ to $T_2$, sequence decoupling schemes are proposed. The decoupling sequence has a wide range of applications in the field of nuclear magnetic resonance, including the spin-echo (SE) sequence [69], which is shown in Fig. 3 (c). The sensitivity of AC measurements using this method is much higher than that of DC measurements [70], and it is currently the highest sensitivity of the NV center magnetometer [71]. However, this approach is equivalent to using a bandpass filter in a specific frequency band, and cannot thus allow for DC or arbitrary frequency measurements. This approach can only be implemented within a specific frequency band of tens of kHz, thus limiting its practicality.

In addition to the SE sequence, more complex dynamical decoupling (DD) sequences exist which achieve longer decoherence time. The most commonly used one is the Carr-Purcell-Meiboom-Gill (CPMG) sequence [72]. At liquid nitrogen temperature (77 K), the CPMG sequence was used to manipulate about 100 NV centers to achieve a $T_2$ decoherence time of 0.6 s [73]. Other possibilities include the periodic DD sequence [74], the Uhrig DD sequence, the concatenated DD sequence [75], the continuous DD [76], and the composite-pulse sequence and rotary-echo (RE) [77]. There are also special methods that can improve the performance of decoupling sequences, such as the quantum interpolation[78] and the quantum phase-locking [79][80]. Furthermore, a classic clock can be employed to eliminate phase randomness between measurements [81][82]. A variety of arbitrary time-varying magnetic field measurement schemes based on NV centers have been proposed [83], including schemes using wavelet transform [84], Walsh reconstruction method[3], dual-channel lock-in [85], and phase estimation algorithms [86]. There are also methods that use the CW-ODMR scheme to measure AC radio frequencies [87].

## 3.2 Comparison of different solutions

### 3.2.1 Comparison of the theoretical sensitivity

Microwave power broadening may cause the PUODMR scheme to be worse than the Ramsey scheme[32]. At the same time, the microwave pulse length in the Ramsey scheme also has an impact on the actual results [88]. For the CWODMR and CWLI scheme, since the spectral is affected by both microwave power broadening and optical power broadening, no theoretical formula exists for the scale factor. There are no existing simulations nor experimental results that compare four schemes together in same condition. Therefore, a simulation analysis was conducted in this review. Eq. (1-16) is used for the simulations, with the selected parameters being consistent with Ref [89]. The optimal conditions and sensitivity of the four methods can be obtained by considering variations in the $T_2^*$, as shown in Fig. 4 (a-d). The sensitivity comparison can be expressed as follows:

$$\eta_{CWLI} \approx 2\eta_{CWUL} \gg \eta_{PUODMR} \approx \eta_{Ramsey} \quad (18)$$

Interestingly, the PUODMR and Ramsey schemes are quite similar, thus indicating that the microwave power has less influence than its own dephasing. The sensitivity and $T_2^*$ for these two pulsed schemes are consistent with the relationship $T_2^{*-0.5} \sim T_2^{*-1}$. $T_2^{*-0.5}$ is considered when $T_2^*$ is much longer than $T_I + T_R$, while $T_2^{*-1}$ is considered in the opposite scenario, as discussed in Ref [32]. The theoretical sensitivity for the two continuous schemes is far worse than that of the pulsed schemes. The gap increases as the $T_2^*$ grows, as the relationship of CW schemes is $T_2^{*-0.38}$. The main reason behind this difference between pulsed and CW schemes lies in the laser-induced broadening effect [90]. The main advantage of the CW schemes over the pulsed schemes is that the photons carry magnetic field information throughout the whole measurement process, while in the pulsed schemes the photons carry information only during detection. The theoretical sensitivity for the CWLI scheme is 2-fold worse than for the direct detection scheme, which indicates the lock-in process lead to loss of information.

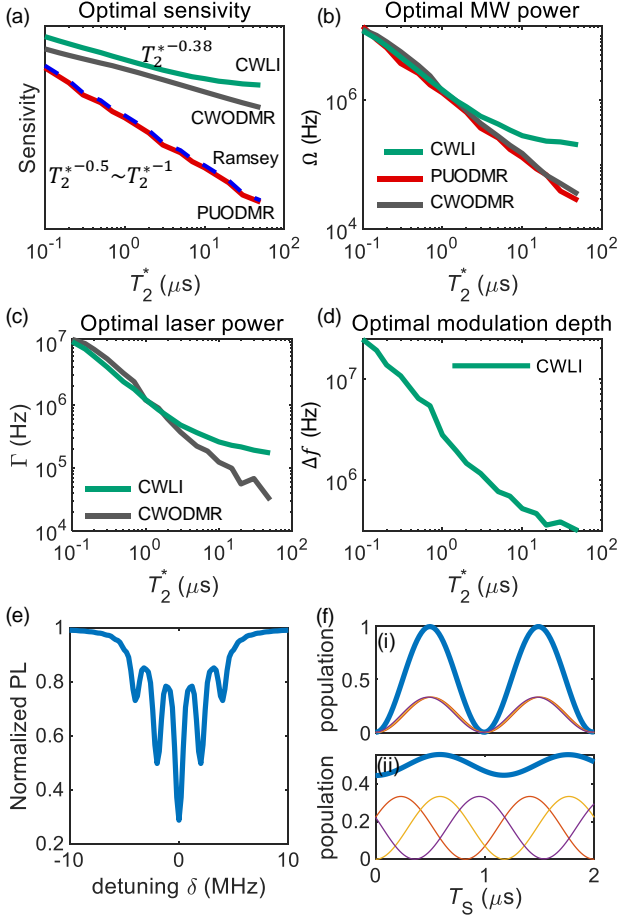

**Fig. 4** (**a**) Relationship between sensitivity and dephasing time for the different methods. (**b**) Relationship between optimal optical pumping rate and dephasing time for the different methods. (**c**) Relationship between optimal microwave power and decoherence time for the different methods. (**d**) Modulation depth of the CWLI method. (**e**) ODMR spectrum under manipulation of three MW frequencies. (**f**) The three independent thin curves represent Ramsey fringes with three hyperfine energy levels, and the thick blue curve represents the total Ramsey fringes. (i) Ramsey fringes of free precession with time $T$ = 3 μs, satisfying Eq. (36). (ii) Ramsey fringes of free precession with time $T$ = 5, which does not satisfy Eq. (36).

### 3.2.2 Comparison of technical noise effects of different schemes

In the actual measurement process, neither equipment nor technology is perfect. Therefore, technical noise will always be present, as shown in Fig. 5 (a). The technical noise terms can be divided into two categories. The first category includes noise terms whose transfer coefficient is independent of the schemes and parameters of the diamond magnetometer, include **magnetic field noise**, **microwave frequency noise**, and **temperature noise**.

**Magnetic field noise.** Since the NV center magnetometer usually requires a device generating a bias magnetic field, the magnetic noise can be divided into two parts, the magnetic noise of the device and the magnetic noise of the environment:

$$\Delta B = \Delta B_{environment} + \Delta B_{circuit} \quad (19)$$

**Microwave frequency noise.** The microwave frequency noise can be equivalent to the magnetic field noise through the electron gyromagnetic ratio, according to:

$$\Delta B = \frac{1}{\gamma_e}\Delta f \quad (20)$$

**Temperature noise.** According to the introduction to Section 2, the temperature fluctuation $\Delta T$ affects the $D_{gs}$, similarly to the microwave frequency noise, its scale factor can be derived as $k_T = \frac{dD_{gs}}{\mu_e dT} = $ 260 nT/K[44], and the corresponding noise is,

$$\Delta B = k_T \Delta T \quad (21)$$

This type of noise is the same for all schemes, hence, no additional comparison is required. The second category includes noise terms whose transfer coefficient is related to the schemes and parameters of the diamond NV center magnetometer, including **detection noise**, **laser power noise**, and **microwave power noise**.

**Detection noise.** The detection noise refers to the noise of the photodetector and its standard deviation is defined as $\Delta I$ or $\Delta \mathcal{F}$. The relative detected fluorescence intensity is defined as: $R_I = I/I_{avg} = \mathcal{F}/\mathcal{F}_{avg}$. The relationship between the standard deviation of the relative detected fluorescence $\Delta R_I$ and the standard deviation of the magnetic field caused by it, is expressed as:

$$\Delta B = {dB}/{dR_I} \Delta R_I \quad (22)$$

**Laser power noise.** Since the laser excitation at room temperature is a non-resonant excitation, the laser or light-emitting diode (LED) wavelength noise can be neglected in this case. Only the power noise needs to be considered. Here, the noise is defined as $\Delta \Gamma$, and the relative optical power is defined as $R_\Gamma = \Gamma/\Gamma_{avg}$. The relative detected fluorescence fluctuations and the relative optical power fluctuations have approximately the same effect on the fluctuation of the magnetic field measurement value[89][91], thus the transfer coefficient is:

$$\frac{dB}{dR_\Gamma} \approx \frac{dB}{dR_I} = k_R \quad (23)$$

**Microwave power noise.** The inaccuracy $\Delta \Omega$ associated to the microwave power can be thought of as equivalent to changing the microwave flip angle, that is, $\frac{\Delta \theta}{\frac{\pi}{2}} \approx \frac{\Delta \Omega}{\Omega}$. By defining the relative microwave power as $R_\Omega = \Omega/\Omega_{avg}$, the corresponding noise is,

$$\Delta B = {dB}/{dR_\Omega} \Delta R_\Omega \quad (24)$$

Since the transfer coefficient of the second category noise is related to the measurement scheme and system parameters, we verify the transfer coefficients of the four schemes through simulation, For one scheme, the impact of these three noise sources on the results is same, thus we can use $k_R$ to represent all there transfer coefficients. The transfer coefficients of the two PU schemes are consistent, while the transfer coefficient of CWODMR is much smaller than the PU method due to a smaller $\frac{dB}{dI}$. The CWLI scheme benefits from the noise suppression of phase sensitive detection and its sensitivity to the noise is more than four orders of magnitude less when compared with the other schemes, as shown in Fig. 5 (b).

For pulse-type schemes, there are two more types of noise: **microwave pulse time noise** and **laser pulse time noise**.

**Microwave pulse time noise.** The effect of an inaccurate microwave pulse length $\Delta t_{MW}$ has similar characteristics as the microwave power noise. It follows that,

$$\frac{\Delta \theta}{\frac{\pi}{2}} \approx \frac{\Delta t_{MW}}{t_{MW}} \quad (25)$$

**Laser pulse time noise.** The inaccuracy of the laser pulse time $\Delta t_{Laser}$ affects both the polarization pulse and the detection pulse. Since polarization is an exponential process[92], the inaccurate pulse time of the polarized light has little effect on the system, and only the inaccurate detection of the pulse time is considered. the laser pulse time noise and the laser power noise can be considered to have the similar effect on the magnetic field. This yields the same as in Eq. (23).

**Scale factor noise.** Laser power noise, microwave power noise, microwave impulse noise, and microwave frequency noise all affect the scale factor, and in turn result in the scale factor noise. It can be calculated that the impact of the scale factor noise, generated by all noises listed above, on the magnetic measurement result is much smaller than the direct impact of those noises themselves. Thus, the scale factor noise can be safely neglected.

**Summary of technical noise.** By adding together the technical noise terms and the photon shot noise, and by taking into account the additive nature of the variance, the total noise becomes:

$$(\Delta B)^2 = \Delta_{shot\ noise}^2 + k_R^2((\Delta R_I)^2 + (\Delta R_\Gamma)^2 + (\Delta R_{Tlaser})^2)$$
$$+ \frac{1}{\gamma_e^2}(\Delta f)^2 + k_\Omega^2((\Delta \theta_\Omega)^2 + (\Delta \theta_{Tmw})^2)$$
$$+ (\Delta B_{environment} + \Delta B_{circuit})^2 + k_T^2(\Delta T)^2 \quad (26)$$

This result neglects the coupling term between noise and the scale factor term as they are assumed to be neglected compared with other noise terms. In this way, the sensitivity formula used in the actual experimental system is finally obtained.

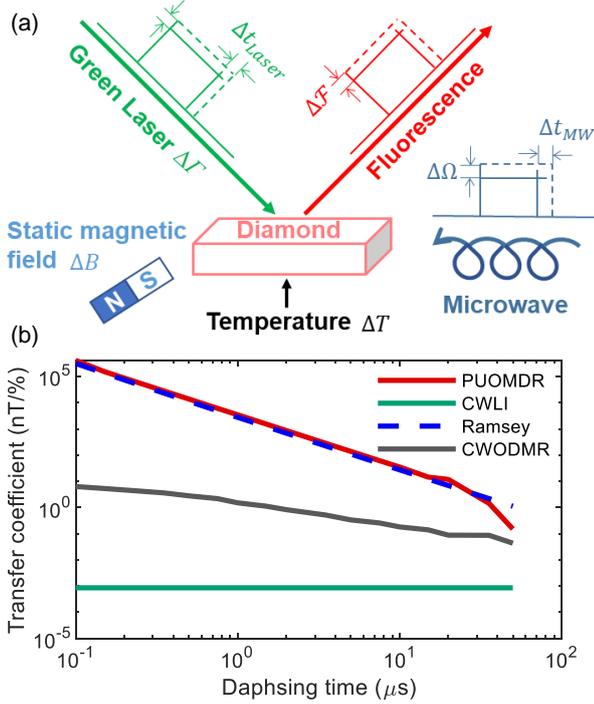

**Fig. 5** (**a**) Schematic illustration of the main parameters of the diamond NV center magnetometer with corresponding noise terms. For the optical spin readout scheme, a laser excitation is required, which introduces the laser power noise $\Delta \Gamma$. Furthermore, the reading noise of the fluorescence detector is $\Delta \mathcal{F}$, the microwave power noise is $\Delta \Omega$, the magnetic field noise of the static magnetic field device is $\Delta B$, the noise caused by temperature fluctuations is $\Delta T$, the pulse time noise is $\Delta t_{Laser}$, while $\Delta t_{MW}$ represents a noise term unique to the pulse type. (**b**) The transfer coefficient $k_R$ of different methods varies with the dephasing time.

### 3.2.3 Removal of the peak splitting effect caused by the $^{14}$N nuclear spin

Due to the hyperfine interaction between the $^{14}$N nuclear spin and the electron spin, the energy levels are split into three. This results in a √3-fold reduction in sensitivity. Therefore, proposing a method to overcome this issue would be beneficial. One solution consists in using the nuclear spin polarization method to polarize the nuclear spin to the desired state. According to the analysis of the hyperfine energy levels in Section 2.2, there are several different ways to achieve nuclear spin polarization, such as the ESLAC method[42], which requires a strong magnetic field and limits application as magnetometers. Another approach involves using microwave and radio frequency to achieve nuclear spin polarization, as discussed in Ref [93][94][95]. This method takes long time to achieve polarization, making it is also not suitable for magnetometer applications.

Compared to polarizing three hyperfine energy levels to one energy level, a more realistic solution would be to use three hyperfine energy levels for simultaneous measurements, as illustrated in in Ref [68][96]. This approach can only be used for the schemes of CWLI and CWODMR. For the PUOMDR scheme, an approach similar to that used in the CWODMR can be proposed in this review, which uses microwaves with three frequencies to scan three peaks simultaneously. Simulations were performed to verify the practicability of this approach. The results of these simulations are shown in Fig. 4 (e) and demonstrate the feasibility of the method.

For the Ramsey scheme, since the optimal $\delta$ occurs at the position that satisfies Eq. (5), in order to achieve the simultaneous use of three hyperfine energy levels, it becomes only necessary to consider the Ramsey free precession time $T$, such that the following relationship holds:

$$(\delta \pm A_\parallel)T_S = \frac{\pi}{2} + j\pi + 2\pi i, i = 0, \pm 1, \pm 2, \cdots$$
$$\delta T_S = \frac{\pi}{2} + j\pi, j = 0, \pm 1, \pm 2, \cdots \quad (27)$$

Due to the limited decoherence time, $i$ cannot be very large. By taking $i = \pm 1$, then $T_S = \frac{2\pi}{A_\parallel} \approx 3$ μs. This $T_2^*$ is attainable, thus this approach is also feasible. The simulation results shown in Fig. 4 (f) confirms this proposal.

### 3.2.4 Summary

According to the analysis, the Ramsey and PUODMR schemes have the highest theoretical sensitivity, but are more sensitive to various technical noises. Therefore, in these two schemes it is difficult to achieve their high theoretical sensitivity limit. Considering the actual benefits of the lock-in detection, most practical magnetometer solutions use CWLI scheme. The highest index of the existing DC magnetic field measurement sensitivity is achieved using the CWLI scheme. Furthermore, combined with closed-loop control, a large measurement range can be achieved [66]. The technical noise value related to the special equipment is evaluated in the section 4.6.2. In addition, the microwave area uniformity and laser spot uniformity lead to a further reduction in sensitivity, which will be also discussed in the key technical part.

# 4 Key technologies

The main experimental devices needed to achieve the diamond NV center magnetic measurement are as follows. A diamond sample with NV center; A light source (generally a laser or LED generating 532 nm green light) to achieve spin polarization and manipulation. Due to the high optical power density required for effective polarization and detection, a focusing device (such as an objective lens) is typically used to converge the beam; A microwave generating device and a microwave antenna are required to achieve spin manipulation; An external magnetic field device, such as a Helmholtz coil or a permanent magnet, is required to achieve the degeneracy of $m_s = \pm 1$. A fluorescence detection device, such as a photodiode or a charge-coupled device (CCD) camera, is also required.

Furthermore, a laser pulse and a microwave pulse generating device are necessary for the pulsed schemes. Typically, pulsed LED or acousto-optic modulator (AOM) are used to generate pulsed light, while microwave switches are employed to generate microwave pulses. The equipment listed above is required to implement the general measurement method. More specific schemes, including the microwave-free scheme and the zero magnetic field scheme, may not need all these devices, or may require additional equipment. These special cases are specifically introduced in Section 4.4 and 4.5.

Several of the experimental devices listed above are commercially mature devices, including the 532 nm laser or LED light source, the avalanche photodiode (APD) or photodiode (PD), the magnetic field source, and the microwave generator. Here, only the technologies that require special development for the NV center magnetometer are discussed as key technologies. These mainly include four areas: sample preparation, microwave feed technology, fluorescence collection, and laser excitation technology, which are discussed in the following Sections.

## 4.1 Diamond sample preparation

According to the sensitivity formulas, the sensitivities of all schemes are inversely proportional to the dephasing time of the sample and to the number of the involved NV centers $N$. Thus, a long dephasing time $T_2^*$ and a high concentration of NV centers enhance the sensitivity, which are characteristics of the sample and are determined by the diamond preparation process.

Because of its uncontrollable size, impurity concentration, and output, natural diamond is rarely used for related research on diamond NV center magnetometry. The diamond used in research is mostly synthetic diamond. According to different preparation methods, there are two dominant types of artificial diamond: chemical vapor deposition (CVD) diamond and high temperature and high pressure (HTHP) diamond. In the CVD approach, one can control the concentration of impurities present in the diamond; this is currently the most widely used approach to NV center diamond preparation[97]. Under certain conditions, the NV center generated in the CVD diamond is directional[98].

The conventional HTHP diamond preparation method requires iron-cobalt-nickel catalysts and has high concentration of N, which results in a larger number of magnetic impurities and a shorter dephasing time[99][100]. Thus, the conventional method is not suitable for NV center sensing applications. However, with new technology, several HTHP samples can achieve the same impurity concentration as the CVD samples, with lower stress inhomogeneity by improving the process[101], thus resulting in overall better performance. Since the $T_2^*$ decoherence time is greatly affected by the $^{13}C$ nuclear spin, $^{12}C$ pure diamond samples have also been extensively studied. Furthermore, diamond nanocrystals prepared for the application of nano-probe magnetometers also exist [102].

Table 1 Summary of diamond sample parameters.

| | Type | $T_2^*$ (ns) | $T_2$ (SE) (ns) | $T_2$ (CPMG) (ns) | $T_1$ | Temperature | $^{13}C$ | concentration (ppm) |
|---|---|---|---|---|---|---|---|---|
| V. M. Acosta et al. [99] | HPHT | 118 | | | | RT | Normal | 16 |
| | CVD | 291 | | | | RT | Normal | 0.012 |
| T. Wolf [71] H.Zheng [162] | HPHT | $11 \times 10^3$? | | | | RT | 0.03 | 0.3 |
| G. Balasubramanian [240] | CVD | $18.2 \times 10^3$ | $1.8 \times 10^6$ | | | RT | 0.3% | Single |
| E.D. Herbschleb [241] | Ntype | $1.5 \times 10^6$ | $2.4 \times 10^6$ | $3.9 \times 10^6$ | 7.4 ms | RT | 0.002% | Single? |
| N. Bar-Gill [73] | CVD | | | $6 \times 10^8$ | >10 s | 77 K | 0.01% | 0.00001 |
| | CVD | | | $5 \times 10^7$ | >10 s | 77 K | 0.01% | 0.01 |
| H. Clevenson [234] | CVD? | 1 μs | | | | RT | Normal | 0.1 |
| P. L. Stanwix [242] | CVD | | $631 \times 10^3$ | | | RT | Normal | 0.00015 |
| | CVD | | $282 \times 10^3$ | | | RT | Normal | 0.0005 |
| J S Hodges [102] | CVD | | $92.6 \times 10^3$ | | | | | Single |
| Balasubramanian[243] | CVD | 10 μs | | | | RT | 0.01% | 3-4 |

Table 2 Summary of microwave antenna parameters.

| | Scheme | Generated magnetic field strength | XY uniform area | Z uniform length | Bandwidth | S11 | Q |
|---|---|---|---|---|---|---|---|
| C. Zhang [100] | Multiple wires | | | | | | |
| K. Bayat [230] | Double loop antenna | 5.6 G @ 0.5 W | $0.95 \times 1.2$ mm² | - | 40 MHz | -25 dB | 70-120 |
| N. Zhang [238] | inset-fed rectangle microstrip patch antenna | 0.8 G @ 1 W | $1 \times 1$ mm² | 0.5 mm? | 60 MHz | -20 dB | - |
| Y.Masuyama [231] | coplanar waveguide resonator | 3.7 G @ ? | $>2 \times 2$ mm² | | Small | - | - |
| K. Sasaki[232] | large-area microwave antenna | 3 G @ 1 W | 1 mm² | >0.4 m | 400 MHz | | 7 |
| W. Jia [239] | coplanar waveguides | 6-10 G @ 1 W | small | | 15.8 GHz | ~ -20 dB | |
| P. Kapitanova[112] | dielectric resonator antenna | 2.8 G @ 5.2 W | Diameter 2.4 mm | 1.7 mm | ~ 50 MHz | | 0.1 |

The most common preparation process of the NV center involves electron irradiation, or ion irradiations, to generate vacancy on the C sites, followed by subsequent annealing at temperatures in the range of 800°C to 1000°C [103], which causes migration of vacancies in the diamond crystal and combine with substitutional N atoms. If the N concentration is insufficient, additional N atoms in diamond can be artificially introduced by $N^+$ ion irradiation [104]. In the preparation process, the indexes that require careful consideration are: the concentration of NV centers, the concentration of the N atoms, the concentration of the impurity color centers, the stress uniformity, the proportion of $^{13}C$ atoms (naturally 1.1%), and the size of the sample. There are numerous studies and literature reviews [105][106][107] on the details of the diamond preparation process, thus it is not necessary to discuss them here further. The parameters for different samples are summarized in Table 1. 1 ppm corresponds to $1.76 \times 10^{17}$ cm$^{-3}$ in the sample. It should be noted that although the polycrystalline CVD diamond in the table can also reach a high performance, it is difficult to use it in magnetic field measurements due to the uncertainty of the internal axial direction[108].

## 4.2 Microwave antenna

After the sample is prepared, the dephasing time and the NV center concentration are fixed. In order to further improve the sensitivity by increasing $N$, a larger participating volume is necessary. This requires simultaneous improvement of the effective control volume for the microwave antenna, the laser excitation, and the fluorescence detection technologies. The microwave antenna is discussed first. The main indicators for evaluating microwave control include microwave feed efficiency S11, control area and uniformity within the area, and microwave frequency bandwidth.

In the past, one or two straight wires were used as microwave antennas, as shown in Fig. 6 (a-b). The main advantage of this method is the simplicity of the structure, while the biggest disadvantage is that the uniform area is small. Another common method is the loop antenna, which can greatly improve the microwave uniform area, but has the intrinsic disadvantage of a limited frequency bandwidth, as shown in Fig. 6 (c).

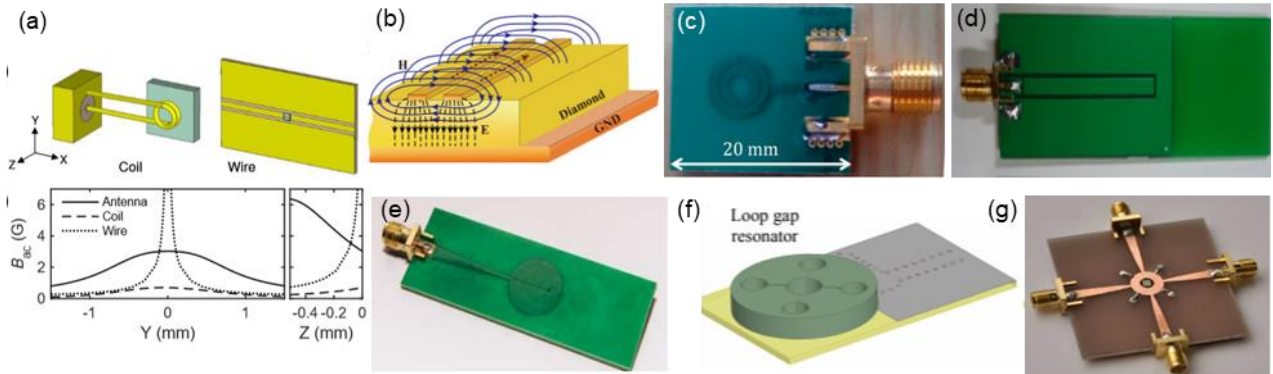

**Fig. 6** (**a**) Single wire scheme and coil scheme [229]. (**b**) Two-wire scheme [100]. (**c**) Double loop antenna [230]. (**d**) Coplanar waveguide resonator [231]. (**e**) Planar ring antenna [232]. (**f**) S-band tunable loop gap resonator [233]. (**g**) Circular polarization antenna [109].

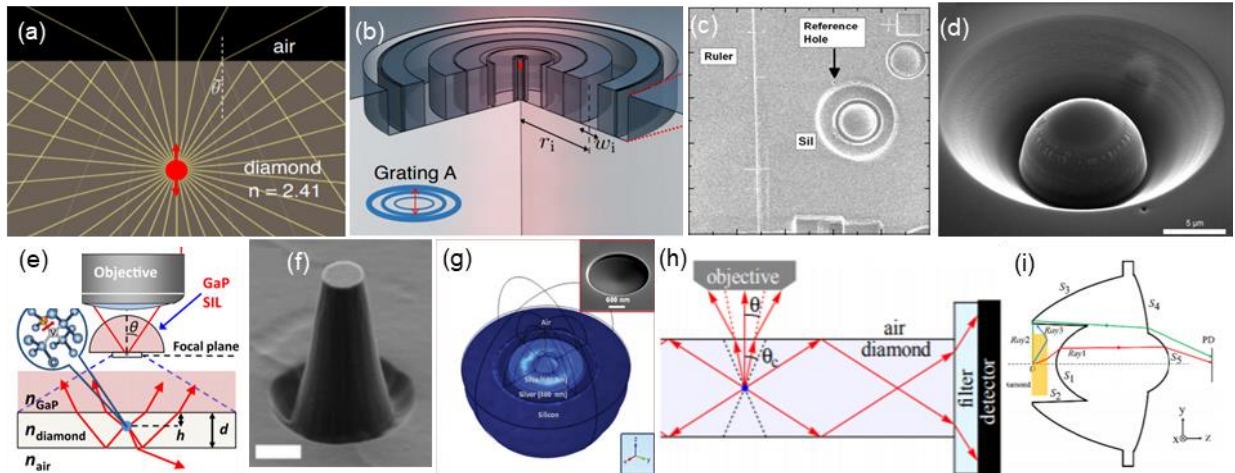

**Fig. 7** (**a**) Schematic diagram of the total reflection. (**b**) Chirped circular dielectric gratings [119]. (**c-e**) Different solid lens SIL schemes [115][235] [118]. (**f**) Optical waveguide schemes [114]. (**g**) Optical waveguide with a nanodiamond in the middle [236]. (**h**) Side collection scheme [124]. (**i**) TIR lens scheme [125].

Specially designed antennas can also be designed, such as the inset-fed rectangle microstrip patch antenna, the strip antenna coplanar waveguide resonator, and the tunable loop gap resonator, as shown in Fig. 6 (d-f). In the absence of an external magnetic field, the $m_s = +1$ and $m_s = −1$ energy levels are degenerate, so the linearly polarized microwave control field cannot resolve the two sublevels. Therefore, in order to achieve the control in this condition, circularly polarized antennas have also been proposed [109][110][111], as shown in Fig. 6 (g). It should be noted that since the NV axis has four different orientations, for diamond samples with no preference on the NV axis orientation, even circularly polarized antennas cannot be directly used for measurements without a special experiment design. Special-purpose antennas have also been implemented, which are used to explore applications under particular circumstances, such as dielectric resonator antennas can achieve greater uniformity in the z-axis [112] and the 60-90 GHz microwave antennas [113]. The technical indicators of the existing microwave antenna solutions are summarized in Table 2.

### 4.3 Fluorescence collection

Fluorescence is typically collected using a confocal light path and objective lens. A major issue for this approach is that, for a bulk diamond, the refractive index of the diamond is 2.4 which leads to a small total reflection angle (24.6°). As a result, the collected fluorescence can be seriously insufficient, as shown in Fig. 7 (a). Therefore, various different methods have been proposed to improve the fluorescence collection efficiency. These methods can be broadly divided into two categories. The first is designed for collecting $nm^3$ to $\mu m^3$ volume, which is mainly single NV center situation. The surface of the diamond around the NV center is usually changed to increase the light extraction efficiency, through approaches that include an optical waveguide [114], a solid immersion lens (SIL) [115][116][117], a dielectric antenna [118], chirped circular dielectric gratings [119], as shown in Fig. 7 (b), circular bullseye gratings [120], plasmonic cavities and gratings [121], all-dielectric nanoantennas [122], waveguide-coupled cavities [123], and nanodiamond placed in the optical waveguide. This method uses an objective lens for collection. as shown in Fig. 7 (g). The second approach is the program for collecting whole diamond bulk, which is mainly for NV center ensemble, such as using a commercial compound parabolic concentrator CPC [71], which achieves a one-sided collection, thus the frontal and side fluorescence cannot be collected simultaneously. The side collection scheme [124] achieves side fluorescence collection, and the frontal fluorescence cannot be effectively collected, as shown in Fig. 7 (h). The total reflection lens (TIR) [125] scheme can achieve fluorescence collection on five faces, as shown in Fig. 7 (i). An antireflection coating [126] is used to reduce the effect of the diamond high refractive index.

### 4.4 Light excitation

There are few studies investigating how to increase the light excitation volume as a separate topic, but the excitation volume is considered as part of the research in many NV center magnetometry studies. The laser excitation includes mainly two indicators: the excitation volume and the excitation optical power density. According to the above analysis, increasing the excitation volume is equivalent to increasing the number of participating NV centers. According to Ref [127], increasing the excitation optical power density is conducive to increasing the sensitivity. Here, it can be considered that the increase in power density corresponds to $C\sqrt{I_{avg}}$ in Eq. (9). The most common excitation scheme is the confocal method. The excitation and detection volumes of this scheme are basically identical, and the laser spot converging by the objective lens leads to high optical power density. However, the excitation area is small. Furthermore, due to the spatial selectivity of the confocal scheme, the diamond that can be collected is also small in the longitudinal direction, causing the total volume to be small.

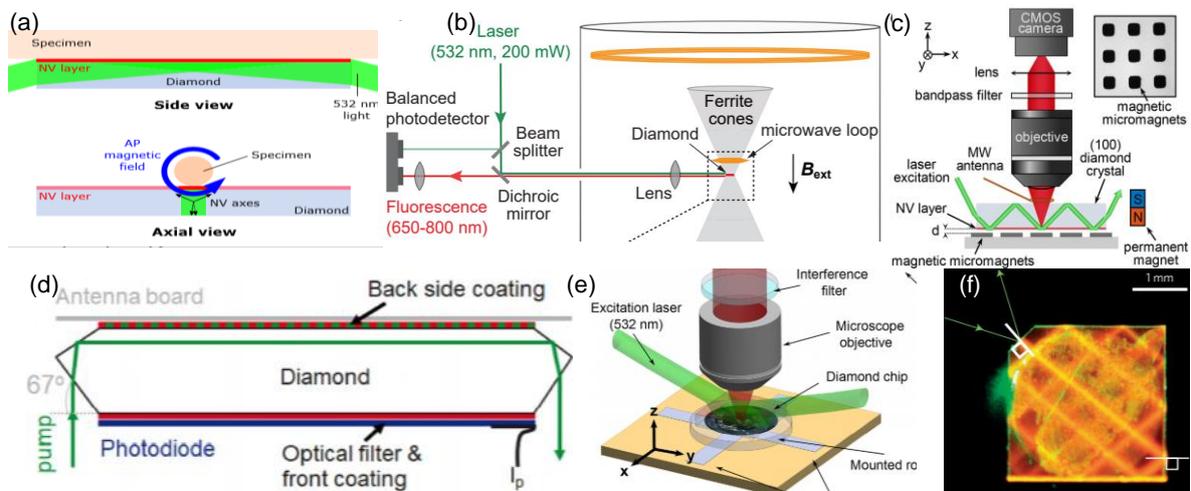

**Fig. 8** (**a**) Side-incident laser solution, the NV center is concentrated on the diamond surface. (**b**) With the addition of ferrite flux, the front side cannot be illuminated, so the side solution is adopted (diamond magnetometer enhanced by ferrite flux concentrators). (**c**) Side-oblique laser incidence [198]. (**d**) Side incidence using the Borst angle [169]. (**e**) Oblique incidence of diamond from the front in the magnetic microscope scheme [200]. (**f**) Light-trapping diamond waveguide approach, which uses multiple reflections inside the diamond to increase the measurement volume through total reflection effect and side incidence [234].

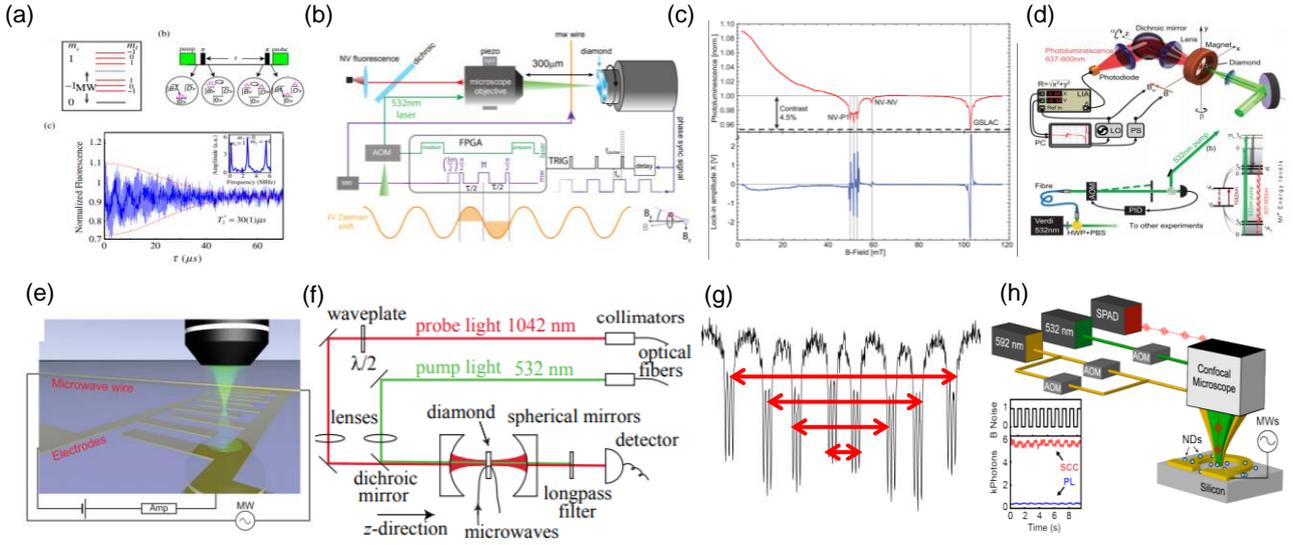

**Fig. 9** (**a**) Quantum beat scheme. (**b**) Magnetic field measurement to achieve $T_2$ decoherence time by rotation. (**c**) LAC principle. (**d**) Experimental setup of microwave-free scheme. (**e**) Schematic diagram of the photocurrent spin detection device [152]. (**f**) Magnetic measurement using the singlet resonance absorption scheme [159]. (**g**) ODMR with four different direction NV center. (**h**) Device diagram of the valence state measurement [155].

The longitudinal volume can be increased by non-confocal detection to improve the sensitivity. For example, in Ref [71] a 400 mW laser, with a converging spot diameter of 47 μm, was used, resulting in an optical power density of ~10 kW/cm$^2$ and an excitation volume of ~ 8.5 × 10$^{-4}$ mm$^3$. Improvements can be achieved by obliquely illuminating the sides, as in the scheme with the highest sensitivity of DC magnetic field measurement [96]. As shown in Fig. 8 (a), the excitation volume is 5 × 10$^{-3}$ mm$^3$. However, in order to increase the optical power density, the laser power reached 4 W, thus limiting the practical use of this scheme. Numerous other solutions exist for the side incidence method, as shown in Fig.8 (b-f). A few specialist schemes have also been proposed, such as pump absorption detection scheme using a 532 nm laser to achieve a sensitivity of 100 nT/Hz$^{0.5}$ [128]. Furthermore, it was demonstrated that multiple reflections of the laser light can be used to obtain a high excitation efficiency and a sensitivity of 200 pT [129].

### 4.5 Special DC magnetic field measurement methods

According to the principles previously discussed, conventional DC schemes have some shortcomings, including the sensitivity to DC noise (such as temperature), short $T_2^*$ dephasing time. In order to overcome these limitations, several special approaches are proposed. Since these approaches can improve only a certain aspect of the shortcomings, the sensitivity cannot be effectively improved compared to conventional schemes, and may result in other inconveniences. Therefore, these methods are only briefly introduced here.

#### 4.5.1 Special schemes for DC noise suppression

For DC noise, although the CWLI scheme can effectively eliminate some technical noise, this solution cannot eliminate the noise that directly affects the Hamiltonian, like the temperature noise[130]. To solve this issue, innovative solutions such as quantum beat scheme have been proposed, which can achieve the suppression of DC noise through the difference of spin $m_s = +1$ and $m_s = -1$, as shown in Fig. 9 (a). Recently, quantum beat schemes have been extensively investigated, in particular in combination with a spin bath drive [131], combining double quantum with a pulse sequence [132] [133][134][135] [32].

Furthermore, several schemes have been proposed that use other parameters to modulate and convert the DC magnetic field to an AC measurement, resulting in the decoherence limit of the magnetic field measurement changing from $T_2^*$ to $T_2$. One schemes is Utilizing rotation to convert a DC magnetic field to an AC magnetic field [136][137], as shown in Fig. 9 (b). This approach has the disadvantage of requiring complex mechanical devices. Another scheme is assisted by nuclear spin[138][139], which can only be applied to a single NV center.

#### 4.5.2 Microwave-free scheme

As discussed previously, an LAC effect occurs under ~500 G magnetic field, which can lead to the NV center fluorescence changing with the magnetic field (where ESLAC occurs), as shown in Fig. 9 (c). According to this principle, a high-performance NV center magnetometer [140][141] [142] has also been designed, as shown in Fig. 9 (d). The advantage of this type of magnetometer is that it does not require a microwave device, thus eliminating microwave noise, but the main issue is that a strong magnetic field is required, which also greatly limits its practicality.

#### 4.5.3 Vector magnetic field measurement

Due to the fixed orientation of the NV center in diamond and the large zero-field splitting, NV center magnetometer is intrinsically directional, thus is a vector magnetometer. The DC and AC magnetometers introduced above are projections of the magnetic field vector onto a single axis and do not use the other three axes. Therefore, if the other three axes can be effectively used, three-dimensional magnetic field vector measurements can be achieved [143][144][145].

#### 4.6.4 Flux concentrator scheme

Referring to the traditional magnetometer scheme such as Hall magentometer, ferromagnetic materials were used to achieve magnetic flux concentrator and attain higher sensitivity [146][147], as shown in Fig. 8 (b). A sensitivity index of ~ 100 fT·Hz$^{-1/2}$ can be achieved [148][149]. The use of a magnetic concentrator is equivalent to amplifying the magnetic field, while not actually improving the magnetic sensitivity of the NV center magnetometer.

### 4.6 Non-fluorescent spin readout method

The sensitivity of the conventional fluorescence readout is limited by the photon shot noise. As discussed in Section 4.3, the photon shot noise is limited by low light collection efficiency and low ODMR contrast.

Several other readout methods exist that might further increase the sensitivity [31].

### 4.6.1 Photocurrent spin readout and charge state spin readout schemes

As charge state conversion is related to the spin state, the photocurrent generated by conversion could also be used for spin readout. This scheme does not require fluorescence detection and only needs electronic equipment, thus having the potential for miniaturization. The relative device is shown in Fig. 9 (e), This scheme has been extensively studied [150][151][152][153][154]. This approach requires laser light for polarization, thus the laser components cannot be replaced. Other studies propose using the valence band for readout, as shown in Fig. 9 (h) [155][156]. Overall, these approaches require further investigations to achieve sufficient sensitivity compared with conventional schemes.

### 4.6.2 Infrared readout solution

The 1042 nm infrared fluorescence corresponding to the singlet level transition can also be used for spin readout[157], as discuss in intrinsic optical dynamics . In this approach, the CWLI infrared absorption scheme is mostly used, and the reflective cavity is employed to increase the light power intensity [158][159][160], as shown in Fig. 9 (f). The sensitivity that can be achieved with this scheme is in the range 22 pT-100 nT.

## 4.7 Equipment and sensitivity

### 4.7.1 The theoretical sensitivity limit

Based on the previous discussion, it is possible to estimate the optimal sensitivity that can be achieved through the existing technology, as well as the sensitivity limit that may be achieved in the future. At present, the optimal sensitivity is 15 pT, while the theoretical sensitivity limit reaches 3 pT. By utilizing the same diamond sample as in Ref [71] and fluorescence collection in Ref [125] and excitation technology can be used to achieve an increase in the measurement volume of at least two orders of magnitude, resulting in a sensitivity index of ~ 20 fT·Hz$^{-1/2}$ that can be achieved through the combination of existing technologies. Under these conditions, the root mean square error (RMSE) of the photon shot noise becomes ~ 0.001%. Furthermore, the theoretical sensitivity limit of the pulsed scheme takes into account the higher sensitivity of the CW scheme, according to Section 3.2.1. Therefore, the theoretical limit of 10 fT·Hz$^{-1/2}$ is a reasonable estimate.

### 4.7.2 The technical sensitivity limit

Although the theoretical sensitivity can reach 10 fT·Hz$^{-1/2}$, the technical noise limits its further optimization. In Section 3.2.2, the influence of the technical noise on the sensitivity was analyzed in terms of the transfer coefficient of the technical noise. Here, magnitude of noise terms related special equipment are discussed and specific solutions to reduce their influence are also addressed.

**Magnetic field noise**. The magnetic field generating device is generally a Helmholtz coil or a permanent magnet. Noise of Helmholtz coil is proportional to the current noise of the power supply, which is in the range of sub-nT and hinders high-sensitivity magnetometer applications. The magnetic noise for a permanent magnet suffers from the current-fluctuation-induced noises[161] and is much smaller than that of a coil. Although the research on the magnetic noise of permanent magnets is lacking, we estimate that it is on the order of sub-pT. The environmental magnetic field noise is linked to the laboratory electromagnetic environment, especially at the power frequency of 50 Hz. Even in the absence of electronic equipment noise, the geomagnetic field itself has noise which follows the 1/f relationship typical of the pink noise.

The highest sensitivity of the commercially available SERF and SQUID magnetometers is in the fT·Hz$^{-1/2}$ order. In order to achieve this level, a magnetic shielding device is needed. This shielding is usually composed of ferrite, permalloy, and other highly permeable materials. The internal bias magnetic fields of shielding must also be sub-nanotesla which is far small than that is needed by the NV center magnetometer. Solutions involves using a circularly polarized MW [162] or optical-selection rules [163] to measure in zero magnetic field, thus the magnetic noise is not representing the main sensitivity limitation of the NV center magnetometer.

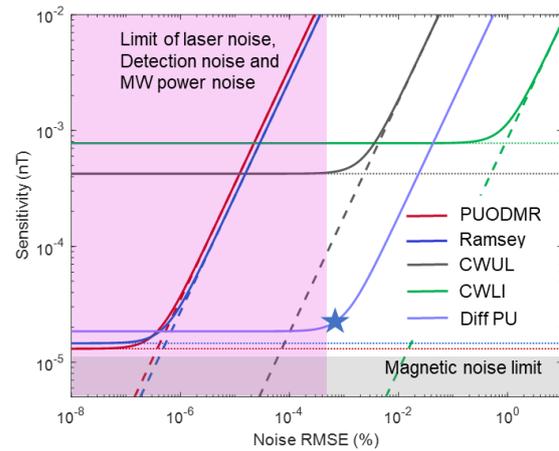

**Fig. 10** The effect of noise on the sensitivity of different schemes is illustrated for a decoherence time of 10 μs and F = 1000. The theoretical sensitivity due to photon shot noise is represented by four dotted lines, and its value is derived from Fig. 4 (a). For the second category noise, the values of the transfer coefficients are from result in Fig. 5 (b). In logarithmic coordinates, the influence of these noise terms on the sensitivity is represented by four dashed lines. Actual sensitivity is shown by the solid line. The RMSE limit of the noise is rarely below 10$^{-3}$%, as indicated by the light pink area. Through the transfer coefficient, noise impact on the actual sensitivity is much greater than the theoretical sensitivity of the PUODMR and Ramsey schemes, but much smaller than the theoretical sensitivity of the CW scheme. After suppressing the technical noise using the differential method, the actual sensitivity of the pulsed scheme becomes the light blue solid line, and the optimal actual sensitivity is 19 fT·Hz$^{-1/2}$, marked by the five-pointed star. Magnetic field noise minimum value is in the order of 10 fT·Hz$^{-1/2}$, as shown by the light gray area.

**Microwave frequency noise**. The highest-quality Agilent microwave sources can achieve a resolution of 0.001 Hz, which means that the corresponding frequency noise should be less than 10 fT·Hz$^{-1/2}$.

**Temperature noise**. Existing equipment is capable to control temperature fluctuations in the order of mK. Furthermore, it should be noted that temperature fluctuations generally occur at low frequency (less than 1 Hz), thus representing a slow drift term that affects more the long-term stability of the magnetometer rather than its sensitivity.

For these first category noise, the conversion scale factor does not change upon changing the theoretical sensitivity limit, thus it becomes the main influencing factor as the sensitivity limit decreases.

**Laser power noise**. The RMSE of laser noise is generally 0.1~1% at low frequencies and ~0.001% on the floor. The laser power noise has a bigger impact on samples with the high $N$ compared to the shot noise. Considering it will be important to increase the number of collected photons through technological improvements, the influence of the laser noise will become greater under this trend.

There are several different methods to reduce the laser power noise. First, the CWLI scheme can largely eliminate the low frequency laser noise. Second, splitting of the laser beam into two beams, one as excitation light and the other as laser power reference light, which is used to differentiate the final result by a balanced detector. Third, for the pulsed scheme, two pulses can be used to differentiate the fluorescence. This differentiation method causes the root of the theory sensitivity to become $2^{0.5}$ times worse[89]. Through these methods, laser noise would be small than shot noise. Additionally, For the CW scheme, a gradient difference scheme can be adopted which can only be used to measure the magnetic gradient[164].

**Detection noise**. Since the sensitivity limit of the conventional detectors is very high, the noise limit of the detector is generally much lower than the laser noise and photon shot noise. Furthermore, the methods to reduce the laser power noise can also apply on the detection noise, thus detection noise will not limit the sensitivity.

**Pulse time noise**. Several works have now demonstrated that the accuracy of the pulse sequence generation device can reach the ps level [165][166][167], while the error associated to the corresponding pulse is less than 0.001%. Thus, the influence of this noise is lower than that of the photon shot noise, and thus it is not considered in this review.

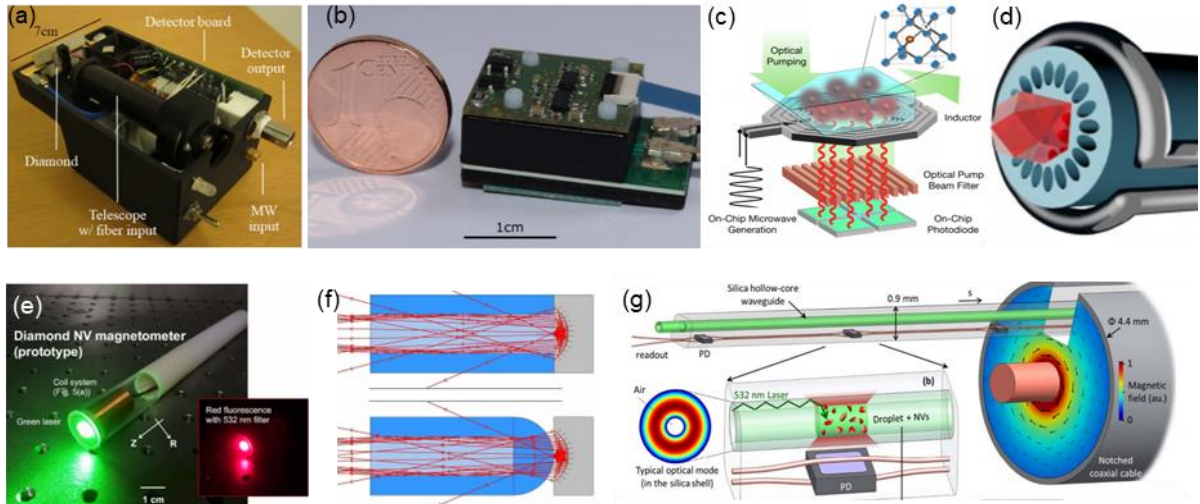

**Fig. 11** (**a**, **b**) Integrated magnetometer scheme [169] [170]. (**c**) CMOS and diamond integrated scheme. [172] (**d**) Fiber optic integrated NV center probe scheme [175]. The laser and fluorescence are transported through the optical fiber, and the microwave line is closely attached to the optical fiber in parallel. (**e**) Prototype of a magnetometer with optical fiber integrated probe. (**f**) Fiber optic probe covering the reflector to improve fluorescence collection efficiency [176]. (**g**) The usage of fiber in the distributed magnetic measurement [183].

**Microwave power noise.** The RMSE of microwave power noise and laser noise at different frequencies are very similar. Both common mode differential method and lock-in scheme can effectively reduce the influence of microwave power noise.

For these second category noise, according to Section 3.2.2, it can be seen that the scale factor of these noise terms decreases upon increasing the dephasing time, thus it becomes crucial to improve the dephasing time of the sample. Finally, we comprehensively consider all technical noise and photon shot noise to get the actual sensitivity. As the result is shown in Fig. 10, the theoretical limit of 10 fT·Hz$^{-1/2}$ is achievable.

# 5 Roadmap of applications

The purpose of the design and manufacture of magnetometers is to use magnetic field measurement applications. In this section, the various application scenarios of special designed NV center magnetometers are shown. These applications represent the future development direction of NV center magnetometers: high spatial resolution and high sensitivity.

## 5.1 The integrated magnetometer

The magnetic measurement system built in a laboratory is typically large and cannot be easily moved [168]. To overcome this practical issue, prototype integrated magnetometers have recently been proposed[169][170]. Fig. 11 (a-b) shows a few hand-held magnetometers with a sensitivity in the order of the nT, mostly limited by the volume of the system [171]. The integration of a complementary metal-oxide semiconductor (CMOS) into a small probe has also been proposed [172][173], as shown in Fig. 11 (c). diamond NV center magnetometer with 200 pT sensitivity is applied to the marine board magnetic anomaly detection[173][174]. This has potential for military use in the anti-submarine applications.

## 5.2 Fiber-combined diamond NV center magnetometer

Schemes using fibers have also been proposed [175][176] [177][178] [179][180] [181]. The general approach of combining an optical fiber with a diamond NV center consists in attaching a small piece of diamond to one end of the fiber, as shown in Fig. 11 (d). This method can provide a small-volume practical magnetometer probe with high integration potential, as shown in Fig. 11 (e). Following this concept, several other improvements have been proposed, such as adding a mirror at the end to improve the collection efficiency of fluorescence and the spatial resolution, as shown in Fig. 11 (f). A tapered optical fiber in endoscope-type configuration has also been proposed to improve excitation and collection efficiency[182]. Nanodiamonds doped into fibers have also been used to perform distributed measurements [183][184] [185], as shown in Fig. 11 (g)

## 5.3 The diamond magnetic microscope imaging scheme

The magnetic field imaging scheme generally uses a diamond NV-rich layer on the surface. The magnetic substance to be tested is placed on the surface of the diamond, and the NV centers at different positions are subjected to different magnetic fields to generate magnetic field imaging. In order to detect the NV centers at different positions, it is necessary to change the optical readout device from PD or APD to a CCD or CMOS camera, as shown in Fig. 12 (a). Early studies were proposed to analyze the feasibility of this scheme [186][187][188][189]. This scheme is characterized by high sensitivity and high spatial resolution. Walsworth's team also made huge progress in this field [190] [191]. The diamond magnetic microscope has applied in many scientific researches, such as measuring microfluidic magnetic spins, imaging system containing magnetic bacteria [192], imaging identification of immunomagnetically labeled cells [193], imaging of malarial hemozoin nanocrystals has also been demonstrated [194], imaging iron biomineralisation in the teeth of the chiton Acanthopleura hirtosa [195], as shown in figure Fig. 12 (b), as well as to detect Mn, Fe, and Gd plasma fluids [196][197].The magneto-optical Kerr effect is combined to further enrich the application scope of this scheme [198]. Furthermore, this scheme has been considered for neuroimaging [199], as shown in Fig. 12 (c). Magnetometer microscopes can be also used to detect micro magnetic ores [148][200]. It is envisaged that magnetic field imaging will find wide applications in several different fields in the future[201].

## 5.4 The AFM probe scheme

By placing a nanodiamond on an AFM probe and scanning the object to be measured, magnetic field imaging with nanometer resolution can be achieved, as shown in Fig. 12 (d). Compared with millimeter resolution gas atomic magnetometer[202], AFM probe scheme brings magnetic field imaging to a whole new level of applicability. The cost of achieving this ultra-high spatial resolution is that the sensitive volume is greatly reduced, resulting in a small number of participating NV centers and the μT-level sensitivity. Several investigations have been performed on the magnetic imaging scheme of the probe[203][204][205][206]. This magnetic field measurement scheme is limited to complex environments and is often not applicable. Instead, $T_1$ or $T_2$ decoherence time measurement is performed. In 2010, a group implemented real-time monitoring of cell membrane ion channels [207], as shown in Fig. 12 (e). The manufacturing method of the probe with an NV center, is based on shallow N ion injection and etching, as shown in Fig. 13 (f) [208].

## 5.5 Nanodiamond for in vivo sensing and cell tracking applications

Thanks to the excellent physicochemical and biological stability of diamond, nanometer diamond particles can be used for quantum measurements of cells and fluorescent labeling[209]. The advantages of nanodiamonds over traditional fluorescent dyeing materials are a good stability without photofading and a very low biological toxicity, thus being an excellent fluorescent labeling material [210], as shown in Fig. 12 (g), such as detection of gadolinium spin labels in an artificial cell membrane [211], as well as the cells temperature [212]. Similar to AFM probe scheme, $T_1$ or $T_2$ decoherence time measurement is commonly performed [111][213][214], as shown in Fig. 12 (h). Nanodiamonds are also used as a medium for drug delivery through chemical group modification. The cell tracking application based on the NV center utilizes its fluorescent properties [215][216][217] or its spin for positioning detection via nuclear magnetic resonance and has been discussed by several reviews [218][219][220][221][222][223][224][225][226][227][228].

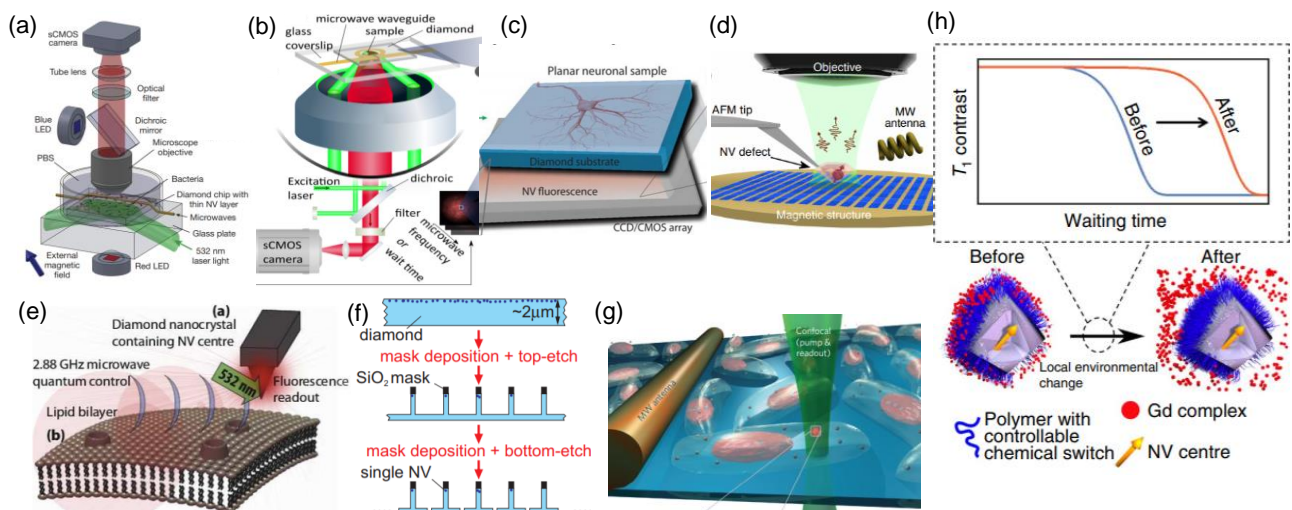

**Fig. 12** (**a,b**) Different magnetic microscope implementation schemes [192] [195]. (**c**) Using magnetic imaging to implement the concept of neural network imaging. (**d**) Schematic diagram for the AFM scheme [204]. (**e**) Measurement of cell membrane ion channels using the probe scheme [207]. (**f**) Schematic diagram of the diamond probe manufacturing process [237]. (**g**) Nanodiamonds used for cell tracking and quantum measurements [209]. (**h**) Quantum measurement scheme for surface-modified nanodiamonds [213].

# 6 Conclusions

In this review, a comprehensive description about diamond NV center magnetometer was demonstrated. Principles and sensitivity of four mainstream DC magnetic measurements schemes were analyzed, and the theoretical sensitivity and vulnerability to technical noise were compared. Assuming the diamond sample performance to be equal, the theoretical sensitivity of the CWLI scheme is the worst but its sensitivity to technical noise is the lowest. The PUODMR scheme and the Ramsey scheme have the highest theoretical sensitivity, while the three non-lock-in schemes are all sensitive to noise. The key technologies required by the diamond NV center magnetometer were discussed: diamond sample preparation, microwave antenna, fluorescence collection, and light excitation. The diamond samples preparation determines the concentration of the NV centers and the decoherence time in diamond. The elimination of $^{13}C$ and the CVD growth technology with impurity content control are consolidated methods to increase the decoherence time and the NV center concentration. Potential methods for future further improvement include the p-type semiconductor diamond technology and the HTHP technology to reduce uneven stress distribution. The main purpose of microwave antenna technology, fluorescence collection, and laser excitation technology is to increase the effective sensitive volume. Among these, the laser excitation technology is the least developed. The impact of the main technical noise affecting the magnetic measurement sensitivity was analyzed, and the technical noise transfer coefficients for the four schemes were compared. Furthermore, the amplitude of each technical noise was also compared for the existing conditions.

Several specialist DC magnetic field measurement schemes were also discussed, such as the microwave-free schemes. In comparison with fluorescent spin readout schemes, non-fluorescent spin readout schemes include valence and electrical spin readout and infrared readout schemes. The signal-to-noise ratio of these two approaches does not currently have many advantages over the fluorescent spin readout scheme.

Through a comprehensive analysis, we predict that the theoretical sensitivity of the diamond NV center magnetometer has the potential to reach the 10 fT·Hz$^{-1/2}$ order without major breakthroughs in the existing technology and theory. As discussed, the crucial issues for NV magnetometer are how to improve dephasing time under complex spin environment and inhomogeneity of zero field splitting and how to combine the existing technologies that are most suitable for improving the sensitivity of the magnetic measurement together. the roadmap for NV magnetometer In the future, diamond preparation may lead to the $T_2^*$ approaching the order of millisecond at low temperatures and lead to a potential of sub-fT·Hz$^{-1/2}$. This will lead to a further increase in the theoretical sensitivity. In the future, a possible technical factor limiting the further improvement of the actual sensitivity might be the magnetic noise. Several specific applications and characteristics of diamond NV center magnetometers were introduced, which demonstrated the roadmap of NV magnetometer application.